\documentclass{iopart}
\usepackage[dvips]{graphics}
\usepackage{graphicx}
\usepackage{amssymb}
\usepackage{color}

\begin{document}

\title[Interactions of solitons with a Gaussian barrier]{Interactions of
solitons with a Gaussian barrier: Splitting and recombination in quasi-1D
and 3D}
\author{J Cuevas$^1$, PG Kevrekidis$^2$, BA Malomed$^3$, P Dyke$^4$ and RG
Hulet$^4$}
\address{$^1$ Grupo de F\'{\i}sica No Lineal. Universidad de Sevilla,
Departamento de F\'{\i}sica Aplicada I, Escuela Polit\'{e}nica
Superior, C/ Virgen de Africa, 7, 41011 Sevilla, Spain}
\address{$^2$ Department of Mathematics and Statistics, University of Massachusetts,
Amherst MA 01003-4515, USA}
\address{$^3$ Department of Physical Electronics, School of Electrical Engineering,
Faculty of Engineering, Tel Aviv University, Tel Aviv 69978, Israel}
\address{$^4$ Department of Physics and Astronomy, Rice University, Houston,
TX 77005, USA}

\ead{\mailto{jcuevas@us.es}}

\begin{abstract}
The interaction of matter-wave solitons with a potential barrier is a
fundamentally important problem, and the splitting and subsequent
recombination of the soliton by the barrier is the essence of soliton
matter-wave interferometry. We demonstrate the three-dimensional (3D)
character of the interactions in the case relevant to ongoing experiments,
where the number of atoms in the soliton is relatively close to the collapse
threshold. The mean-field description is quite accurate, but the proximity
to the collapse makes the use of the 1D Gross-Pitaevskii equation (GPE)
irrelevant. We examine the soliton dynamics in the framework of the
effectively 1D nonpolynomial Schr{\"{o}}dinger equation (NPSE), which admits
the collapse in a modified form, and in parallel we use the full 3D GPE.
Both approaches produce similar results which are, however, very different
from those produced in recent work which used the 1D cubic GPE. Basic
features, produced by the NPSE and the 3D GPE alike, include (a) an increase
in the first reflection coefficient for increasing barrier height and
decreasing atom number; (b) large variation of the secondary
reflection/recombination probability vs.~barrier height; (c) a pronounced
asymmetry in the oscillation amplitudes of the transmitted and reflected
fragments; (d) an enhancement of the transverse excitations as the number of
atoms is increased. We also explore effects produced by variations of the
barrier width and outcomes of the secondary collision upon phase imprinting
on the fragment in one arm of the interferometer.
\end{abstract}


\submitto{\NJP}

\maketitle

\section{Introduction}

Ongoing studies of atomic Bose-Einstein condensates (BECs) have contributed
numerous fundamental insights in a wide range of phenomena~\cite{book1,book2}%
. This may be largely attributed to the precise control afforded by
experiments, and to the existence of accurate, yet quite tractable models
based on the Gross-Pitaevskii equation (GPE)~\cite{rmp,ourbook}. Many of the
theoretical and experimental investigations have strong connections to other
areas, such as condensed matter physics, nonlinear optics,
superconductivity, and superfluidity. Furthermore, ultracold gases trapped
in external potentials may be utilized as quantum simulators of real
materials \cite{emulators}.

Inter-atomic interactions in a BEC enable the examination of a wide variety
of nonlinear effects that have been summarized in a number of reviews and
books such as \cite{ourbook}. Basic types of coherent structures built by
the nonlinearity are bright solitons and soliton complexes \cite%
{chap01:njp2003b,kono,fatk,Torner,extra-reviews}, dark solitons~\cite{djf},
vortices, and vortex lattices~\cite{fetter,us,Japan,fetter2}. Bright
solitons were proposed for matter-wave interferometry \cite{chap01:bright1},
which is itself an active field of research \cite{interferometry}. Recent
work focused on the role of the condensate's effective nonlinearity \cite%
{recent}, as well as on advanced applications \cite{appl}. Solitons may
provide for 100-fold improved sensitivity for atom interferometry, where
their long lifetime, of the order of seconds, may enable precise force
sensing~\cite{markk} and measurement of small magnetic field gradients \cite%
{Rosanov}. Robust bright solitons, appropriate for these applications, have
been created under well-controlled conditions in BECs with attractive \cite%
{chap01:bright1,chap01:bright2,chap01:bright3} and repulsive \cite%
{gap-solitons}\ inter-atomic interactions (in the latter case, these were
gap solitons \cite{kono} supported by the optical-lattice potential). A
topic of particular interest has been the splitting and subsequent
recombination of solitons due to collisions with potential barriers. This
was studied theoretically in detail \cite{splitter,billam1,martin,Lev}, as
summarized in a recent review~\cite{billam2}. However, it is relevant to
stress that the analysis has been thus far carried out for a one-dimensional
(1D) model, i.e., the 1D GPE with cubic nonlinearity
(although some features such as solitary wave collisions have also been
recently explored in a 3D setting~\cite{Parker}). In the same framework,
the quasiparticle-wave duality of solitons and their tunneling through a
potential barrier was recently considered~\cite{Taiwan,Denmark}.

The subject of the present work is the interaction of the incident soliton
with a potential barrier, acting as a beam splitter, and the subsequent
recombination of the split fragments in the 3D setting. The
three-dimensionality strongly affects the results in comparison with the 1D
setting of the above recent studies. Bright matter-wave solitons are
experimentally created in cigar-shaped (nearly-1D) potential traps in which
the self-trapping is driven by the intrinsic attractive nonlinearity solely
in the axial direction. The number of atoms in the soliton, $N$ will be
considered relatively close to the value at the collapse threshold, $N_{c}$,
and the actual shape of the solitons will be three-dimensional in the
considerations that follow. Accordingly, the solitary waves do not reproduce
the conspicuous elongation of the underlying trapping potential, as known
from previous works \cite{chap01:njp2003b,chap01:bright1,chap01:bright2,chap01:bright3}. Thus,
the character of the splitting and subsequent merger may be essentially
three-dimensional too---in particular, transverse modes may (and will, as we
show) be excited.

In the recently studied 1D counterpart of this setting, results of an
earlier work \cite{holmer} were used in \cite{billam1} to predict the
transmission and reflection coefficients and the phase shift resulting from
the interaction with a narrow barrier. A sinusoidal dependence of the
transmission coefficient on the relative phase of two solitary waves
colliding on the barrier was identified. In \cite{martin}, both the primary
splitting of a single solitary wave and the secondary collision of the two
fragments was studied using the truncated Wigner approximation, with
emphasis on the comparison of the mean-field dynamics with the quantum
dynamics. Significant deviations between the two were identified for a
relatively small number of atoms in the solitary wave. A step-wise
(discontinuous) variation of the reflection and transmission coefficients
due to quantum superposition was recently reported \cite{Lev,France,Taiwan}.
The deviation between quantum and mean-field approaches is significant only
for relatively small $N$, becoming negligible for large $N$. In this paper,
we consider $N$ close to collapse, where $N_{c} \simeq 10^5$. For this
reason, quantum effects are not crucially important to us.

Collisions of 1D solitons with an attractive potential well, rather than the
repulsive barrier, have also been studied recently, revealing a fairly
complex phenomenology. The latter features alternating windows of
transmission, reflection, and trapping with abrupt transitions between them
as the potential depth varies~\cite{pantofl1,pantofl2}.

Our analysis is performed in two stages. First, we use the well-known 1D
nonpolynomial Schr{\"{o}}dinger equation (NPSE)~\cite{salas,Luca}, which,
unlike the cubic 1D GPE, captures the 3D phenomenology in some approximation
(in particular, it admits the possibility of collapse, although not in the
same way as the 3D GPE). We will refer to this NPSE setting as a \textit{%
quasi-1D} description to clearly distinguish it from the 1D GPE setting
where collapse is absent. Secondly, we report results obtained from the full
3D GPE. To the best of our knowledge, both equations are used for the first
time in this context. Our main interest is to consider relatively large atom
numbers $N$ ($N=80000\simeq 0.85N_{c}$), for which the phase coherence of
the condensate is maintained and the soliton dynamics are most interesting.
We explore the dependence of the results on $N$, as well as on the height of
the barrier, $E$. 
There exist significant differences between the dynamics described by the
ordinary cubic 1D GPE and predictions of the full 3D equation, as the
excitation of transverse oscillations (the feature which is obviously absent
in the 1D description) becomes prominent for $N\rightarrow N_{c}$. On the
other hand, the quasi-1D NPSE is found to be in reasonable agreement with
the 3D GPE both in regard to its static properties (although the stable
branch of solutions at low $N$ is captured more accurately than the unstable
collapsing one of high $N$) and to the dynamics of soliton-barrier
interactions.

The paper is structured as follows. In Section 2, we introduce the quasi-1D
and 3D models, compute their stationary solutions, and define diagnostics
used to examine the dynamics. In Section 3, we report numerical results in
detail, varying the barrier height and width, atom number, as well as phase
imprinting applied to one of the fragments for simulating the
interferometry. The paper is concluded by Section 4.

\section{Model equations and stationary solutions}

We start by introducing the quasi-1D NPSE for the mean-field wave function, $%
\psi $, of the condensate of $^{7}$Li atoms loaded into a cigar-shaped trap
with axial coordinate $z$,

\begin{equation}
i\hbar \frac{\partial \psi }{\partial t}=-\frac{\hbar ^{2}}{2m}\frac{%
\partial ^{2}\psi }{\partial z^{2}}+V(z)\psi +\frac{1+(3/2)g|\psi |^{2}}{%
\sqrt{1+g|\psi |^{2}}}\psi ,  \label{eq:dyn1D}
\end{equation}%
where $g=2a_{s}/a_{\perp }$, with the typical scattering length $a_{s}=-0.3$
$a_0$, where $a_0$ is the Bohr radius, and transverse trapping radius $%
a_{\perp }=\sqrt{\hbar /(m\omega _{\perp })}=2.25$ $\mathrm{\mu }$m, which
corresponds to a confinement frequency $\omega _{\perp }=2\pi \cdot 290$ Hz
and atomic mass $m=7$ amu. The longitudinal potential $V(z)$ includes a weak
parabolic trap, which we assume has a typical value of the strength, $\omega
_{z}=2\pi \cdot 5.6$ Hz, and the Gaussian barrier of height $E$ and width $%
\varepsilon $, which will be varied below:

\begin{equation}
V(z)=(1/2)m\omega _{z}^{2}z^{2}+E\exp \left( -2z^{2}/\varepsilon ^{2}\right)
.  \label{eq:potential1D}
\end{equation}%
The total number of atoms, given by $N=\int_{-\infty }^{+\infty
}dz\left\vert \psi (z)\right\vert ^{2}$, will also be varied.

The results produced by the NPSE (\ref{eq:dyn1D}) will be compared to those
obtained from the radially symmetric 3D GPE, written in the cylindrical
coordinates, $\left( \rho ,z\right) $ as

\begin{equation}
\fl i\hbar \frac{\partial \psi }{\partial t}=-\frac{\hbar ^{2}}{2m}\left(
\frac{\partial ^{2}\psi }{\partial \rho ^{2}}+\frac{1}{\rho }\frac{\partial
\psi }{\partial \rho }+\frac{\partial ^{2}\psi }{\partial z^{2}}\right)
+V(\rho ,z)\psi +\frac{4\pi \hbar ^{2}}{m}a_{s}\left\vert \psi \right\vert
^{2}\psi ,  \label{eq:dyn3D}
\end{equation}%
with $N=2\pi \int_{0}^{\infty }\rho d\rho \int_{-\infty }^{+\infty
}dz\left\vert \psi \left( \rho ,z\right) \right\vert ^{2}$ and the 3D
potential,

\begin{equation}
V(\rho ,z)=(1/2)m(\omega _{\perp }\rho ^{2}+\omega _{z}^{2}z^{2})+E\exp
\left( -2z^{2}/\varepsilon ^{2}\right) .  \label{eq:potential2D}
\end{equation}%
Equation (\ref{eq:potential2D}) implies that we consider only fundamental
axially symmetric solitons, but not ones with intrinsic axial vorticity,
which are also possible in this setting \cite{Luca}. Axial widths of the
solitons generated by the quasi-1D NPSE and 3D GPE are defined as%
\begin{equation}
\fl W_{\mathrm{1D}}=2\ \sqrt{N^{-1}\int_{-\infty }^{+\infty }dz\,z^{2}|\psi
|^{2}},\ ~W_{\mathrm{3D}}=2\ \sqrt{N^{-1}\ \int_{0}^{\infty }\rho d\rho
\int_{-\infty }^{+\infty }dz\,z^{2}|\psi |^{2}}.  \label{WW}
\end{equation}

\begin{figure}[tbp]
\begin{center}
\begin{tabular}{cc}
\includegraphics[bb=0 0 541 414,width=.45\textwidth]{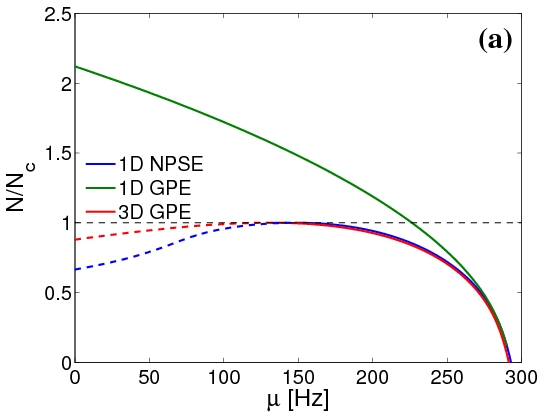} & %
\includegraphics[bb=0 0 541 414,width=.45\textwidth]{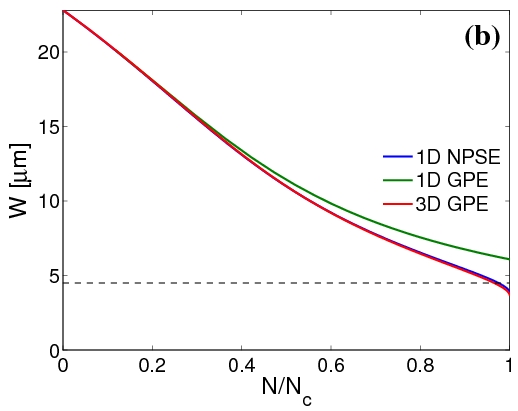}%
\end{tabular}%
\end{center}
\caption{(a) The number of atoms, normalized to its critical value at the
collapse threshold, $N/N_{c}$, versus the chemical potential, for the stable
($\protect\mu >\protect\mu _{c}$) and unstable ($\protect\mu <\protect\mu %
_{c}$) soliton families; $\protect\mu _{c}=147.9$ Hz for the quasi-1D NPSE (%
\protect\ref{eq:dyn1D}), and $\protect\mu _{c}=130.5$ Hz for the 3D GPE (%
\protect\ref{eq:dyn3D}). (b) The axial widths, defined as per (\protect\ref%
{WW}), versus $N/N_{c}$, for the stable soliton family. The dark dashed
horizontal curve in the right panel corresponds to the width of $4.5$ $%
\mathrm{\protect\mu }$m, which is a typical width of the barrier in what
follows. The reference value of chemical potential $\protect\mu $ for
vanishing interaction is given by the transverse oscillator frequency, here $%
\protect\omega _{\perp }/(2\protect\pi )=290$ Hz. The limit value of the
width for the vanishing interaction strength is the confinement length of
the longitudinal trapping potential,
$W=\protect\sqrt{2\hbar /m \omega _{z}}=22.8$ $\mathrm{\protect\mu }$m.
Both panels also
display the comparison with the results produced by the usual 1D
cubic GPE; as expected, that equation is only able to reproduce
results for a relatively small number of atoms, $N\lesssim 0.5N_{c}$.}
\label{fig:norm}
\end{figure}

Stationary solutions with chemical potential $\mu $ were sought for as $\psi
(\mathbf{r},t)=\exp (-i\mu t/\hbar )\psi (\mathbf{r})$. Figure \ref{fig:norm}
compares $N(\mu )$ and $W(N)$ curves for the soliton families, produced by
both the quasi-1D and 3D equations. In these plots, $N$ is normalized to the
critical atom number, $N_{c}$, which initializes the collapse, i.e., the
largest number atoms accessible (by a stable standing wave solution) in the
given setting~\cite{sulem,PhysRep}. Both equations yield nearly the same
value, $N_{c}\approx 94000$. The same plot includes the curves for the
usual 1D cubic GPE, demonstrating that the latter model progressively
deviates from the NPSE and the 3D GPE
for $N/N_{c}\gtrsim 0.5$. Importantly, it is relevant to note that
the solitons of the fomer model have a monotonic dependence of
$N$ on $\mu$ corroborating their dynamical stability for all values
of the chemical potential, contrary to the instability of the latter
two models for $N>N_c$. It is for these reasons that for the
range of atom numbers examined herein, we will restrict our considerations
hereafter to the latter two models.

A straightforward analysis demonstrates that the general condition for the
validity of the NPSE approximation is $N/N_c \simeq N|a|/a_{\perp } < 1$.
Indeed, it is observed in Fig. \ref{fig:norm} that stable solution branches
generated by the NPSE\ and 3D GPE are very close, while the discrepancy is
conspicuous for the unstable branches. In the right panel of Figure \ref%
{fig:norm}, a horizontal line is drawn at $W=4.5$ $\mathrm{\mu }$m, which
corresponds to the typical width of the barrier $\varepsilon $ in the
simulations reported below. It is seen that stable solitons are narrower
than $4.5$ $\mathrm{\mu } $m only at $N>0.98N_{c}$. Consequently, solitons
are typically wider than the considered barrier.

Notice that in what follows, we will keep the trap strength fixed and will
vary the number of atoms $N$. We will be interested in the regime of large $%
N $ (and close to the onset of collapse), where the mean field approximation
above is expected to adequately describe the system. On the contrary, for
cases of tighter confinement or of lower atom numbers, effects of quantum
fluctuations are expected to progressively become more important~\cite%
{martin}.

\section{Numerical simulations}

\subsection{Effect of changing the number of atoms}

We first consider simulations for the solitons launched with zero velocity
at the distance $A=251$ $\mathrm{\mu }$m from the central position, where
the barrier is placed. In the present setting, this amounts to the initial
potential energy equal to $678$ Hz. The barrier is here taken with a fixed
width, $\varepsilon =4.5$ $\mathrm{\mu }$m. The number of atoms in the
soliton and the barrier height $E$ are varied and the outcome of the
soliton-barrier interaction is monitored. The splitting is characterized by
the time-dependent reflection coefficient,

\begin{equation}
R_{\mathrm{1D}}(t)=N^{-1}\int_{-\infty }^{0}dz\,|\psi |^{2},\ R_{\mathrm{3D}%
}(t)=N^{-1}\int_{0}^{\infty }\rho d\rho \int_{-\infty }^{0}dz\,|\psi |^{2},
\label{R}
\end{equation}%
which shows the fraction of atoms remaining in the incident arm of the
interferometer. Below, we focus on $R$ taken at two specific times, $%
R_{1}\equiv R(t=\pi /\omega _{z})$ and $R_{2}\equiv R(t=2\pi /\omega _{z})$.
The former value measures the fraction of atoms that remain in the original
arm after the interaction of the incoming soliton with the barrier, because $%
t=\pi /\omega _{z}$ corresponds to a half-period of oscillation of the
soliton in the weak longitudinal trap, thus ensuring that its first
collision with the barrier (occurring roughly at a quarter of the period) is
completed. The latter value, $R_{2}$, measures the relevant fraction after
the second interaction, i.e., the collision of the returning fragments with
the potential barrier. This occurs at roughly three quarters of the
oscillation period, while $R_{2} $ is measured at a full period (defined in
the absence of the barrier).

\begin{figure}[tbp]
\begin{center}
\begin{tabular}{cc}
\includegraphics[bb=0 0 1200 900,width=.51\textwidth]{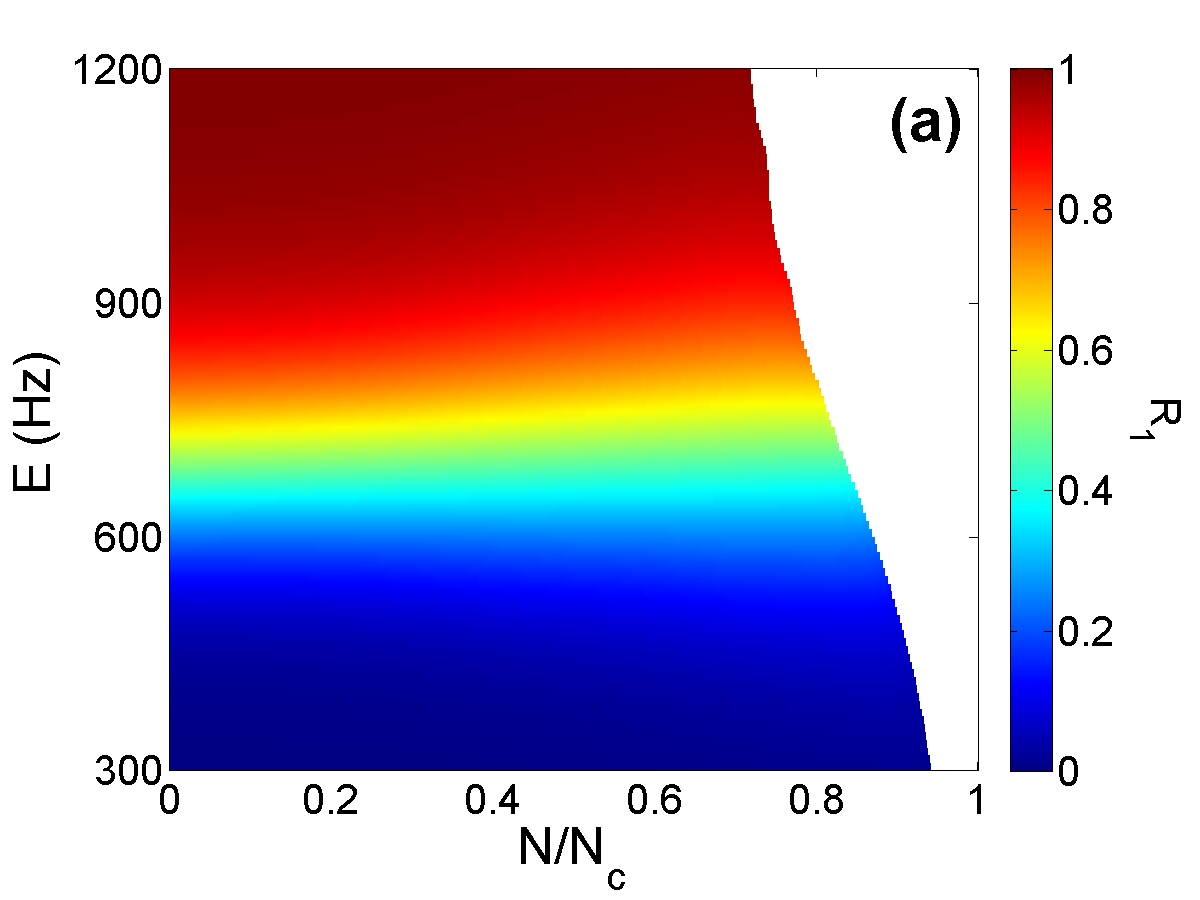} & %
\includegraphics[bb=0 0 1200 900,width=.51\textwidth]{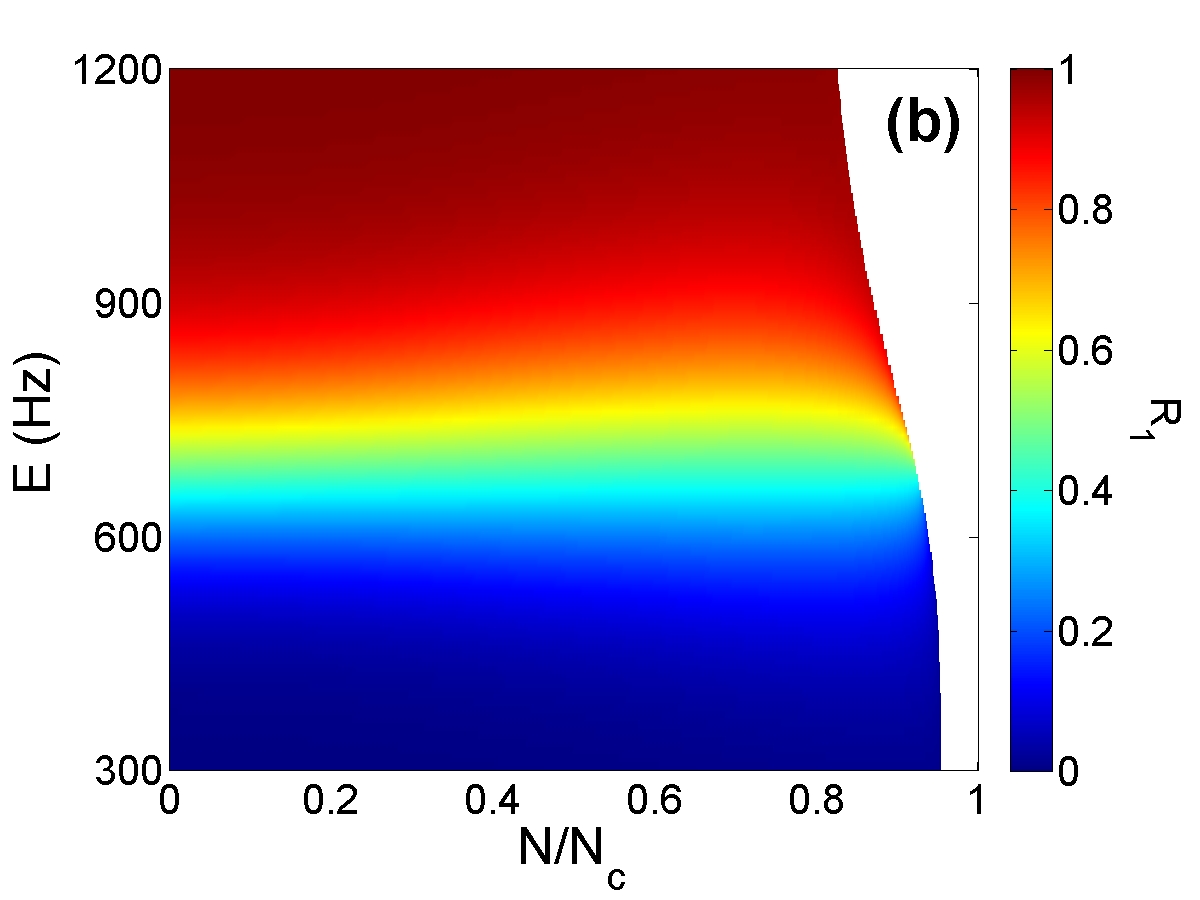} \\
\multicolumn{2}{c}{\includegraphics[bb=0 0 541 414,width=.45\textwidth]{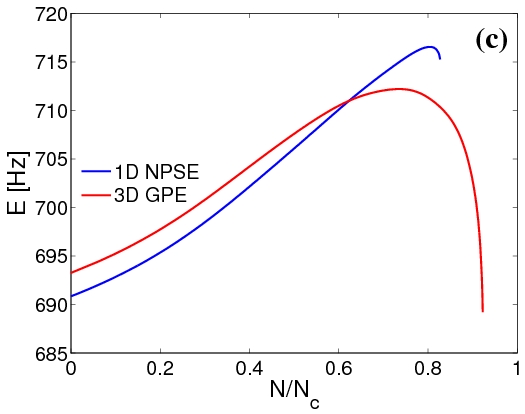}}%
\end{tabular}%
\end{center}
\caption{The first reflection coefficient, $R_{1}$ [see (\protect\ref{R})]
vs.~barrier height, $E$, and the normalized number of atoms, $N/N_{c}$, in
the quasi-1D (a) and 3D (b) settings. Blank regions correspond to
barrier-induced collapse. Panel (c) shows the value of the barrier height at
which equal splitting ($R_{1}=0.5$) is attained, as a function of $N/N_{c}$.
The barrier width is $\protect\varepsilon=4.5$ $\mathrm{\protect\mu}$m.}
\label{fig:R1}
\end{figure}

\begin{figure}[tbp]
\begin{center}
\begin{tabular}{cc}
\includegraphics[bb=0 0 1200 900,width=.51\textwidth]{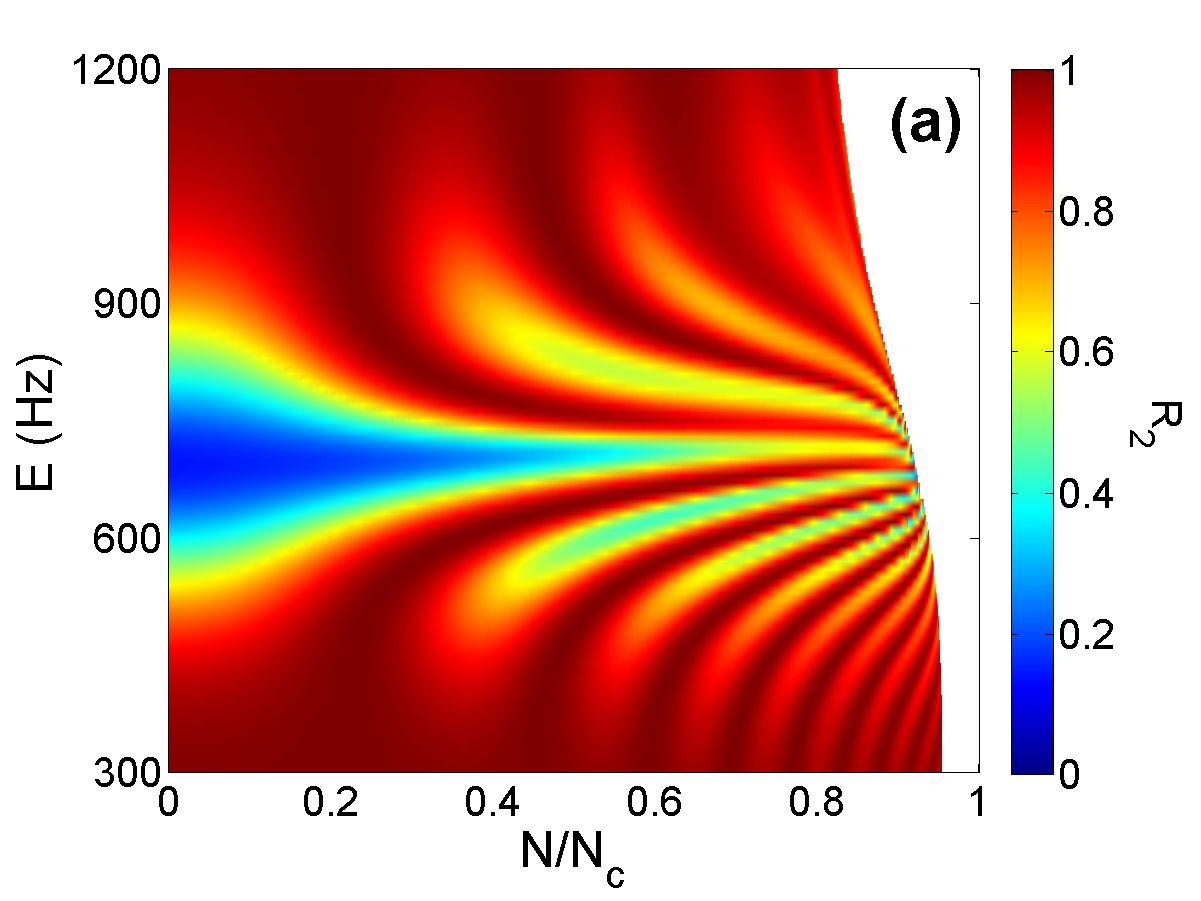} & %
\includegraphics[bb=0 0 541 414,width=.45\textwidth]{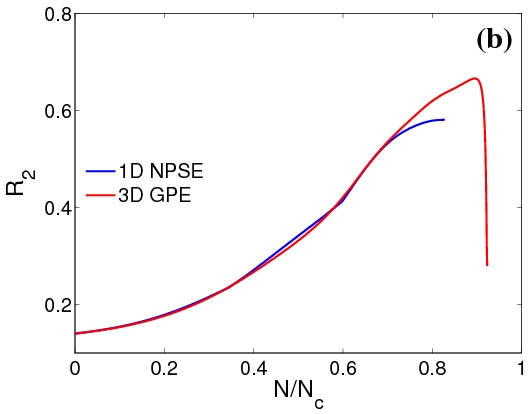} %
\end{tabular}%
\end{center}
\caption{Panel (a) shows the second reflection coefficient, $R_{2}$ [see (%
\protect\ref{R})] vs.~the barrier height, $E$, and the normalized number of
atoms, $N/N_{c}$, in the 3D setting. The barrier-induced collapse occurs in
blank areas. Panel (b) shows the typically increasing trend of $R_2$ as a
function of $N/N_{c}$ when $R_{1}=0.5$. The barrier width is $\protect%
\varepsilon=4.5$ $\mathrm{\protect\mu}$m.}
\label{fig:R2}
\end{figure}

Figures \ref{fig:R1} and \ref{fig:R2} show, respectively, $R_{1}$ and $R_{2}$
versus the number of atoms, $N$, and the barrier height, $E$, with the blank
regions denoting the presence of collapse. Collapse occurs for $N < N_c$ in
these cases due to the interaction with the barrier. It is observed that the
quasi-1D NPSE\ and 3D GPE dynamics show similar trends for the first
reflection. Namely, the reflection increases with the growth of the barrier
height and with decreasing $N$. The second reflection coefficient, $R_{2}$,
presents a more complex functional dependence, which, in the terms of the
quasi-1D NPSE and 3D GPE alike, features an oscillatory variation with $N$
and $E$ for sufficiently large $N$. Nevertheless, we observe good
qualitative and even semi-quantitative agreement between the quasi-1D and
the fully 3D models in this context. This behavior can be attributed to the
number-dependent variation of phases and amplitudes of the fragments
emerging from the first collision, which, upon their recombination, leads to
outcomes ranging from the nearly perfect reflection of the recombined
soliton to its nearly perfect transmission. This is a more complex
manifestation of the feature observed in Ref.~\cite{billam1} and is a
nonlinear effect stemming from the interference of the phases of the two
fragments, becoming progressively more pronounced as $N\rightarrow N_{c}$.
This phenomenon is especially visible for $N/N_c\gtrsim0.3$, leading to a
series of resonant peaks of full reflection.
Another interesting feature is the
presence of finger-like gaps in the quasi-1D setting where collapse occurs
after the second interaction with the barrier, when $N/N_c\gtrsim0.85$.
Those gaps do not appear in the 3D setting (not shown here in detail),
indicating that they originate from an inherent limitation of the NPSE
approximation. This is explained by the fact that in the framework of the
NPSE (\ref{eq:dyn1D}) the soliton collapses when the denominator vanishes,
at the critical value of the peak density (contrary to the genuine collapse
of the fully 3D setting). This difference also justifies the apparently
wider range of parameters (in the $(E,N/N_c)$ parametric plane) leading to
collapse for the quasi-1D problem in comparison to the fully 3D one.
Finally, the right panel of Figure \ref{fig:R2} shows the increase in the
second reflection with increasing $N$, for the case in which $E$ is adjusted
to give 50\% splitting after the first reflection.

While the reflectivity provide a measure of the asymmetries between the
atom-number fractions emerging to the left and right of the barrier, we have
used additional diagnostics to quantify the dynamics. In particular as a
measure of the asymmetry of the reflected and transmitted waveforms, we
define
\begin{equation}
\zeta \equiv \frac{A_{t}-A_{r}}{A_{t}+A_{r}},  \label{zeta}
\end{equation}%
where $A_{r}$ and $A_{t}$ are the oscillation amplitudes within the trap of
the reflected and transmitted fragments, and are both taken to be positive.
The dependence of $\zeta $ on $N/N_{c}$ and $E$ is presented in Figure \ref%
{fig:zeta}, along with its dependence on $N/N_{c}$ when $E$ is adjusted to
give 50\% splitting. An interesting feature is that $\zeta >0$ for every $N$
and $E$, i.e., the transmitted fragment always reaches a higher value of $|z|
$ and, consequently, has a higher kinetic energy, independently of whether
it is the larger or smaller fragment. This can presumably be attributed to
the original direction of motion (and associated momentum) of the incoming
solitary wave. It should also be noted that typically the most pronounced
asymmetries occur for larger values of $N/N_{c}$, appearing to be
predominantly a feature of the 3D nature of the interactions in that case.

\begin{figure}[tbp]
\begin{center}
\begin{tabular}{cc}
\includegraphics[bb=0 0 1200 900,width=.51\textwidth]{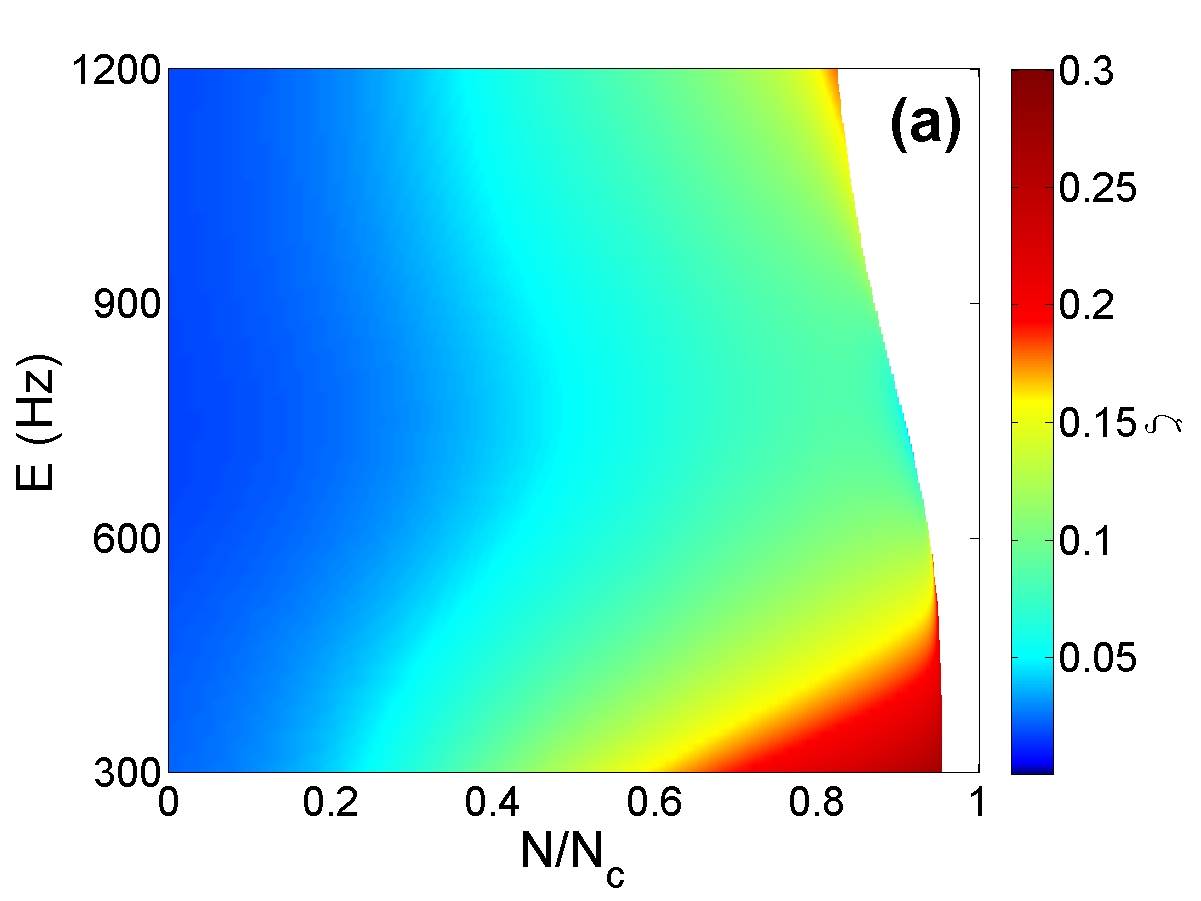} & %
\includegraphics[bb=0 0 541 414,width=.45\textwidth]{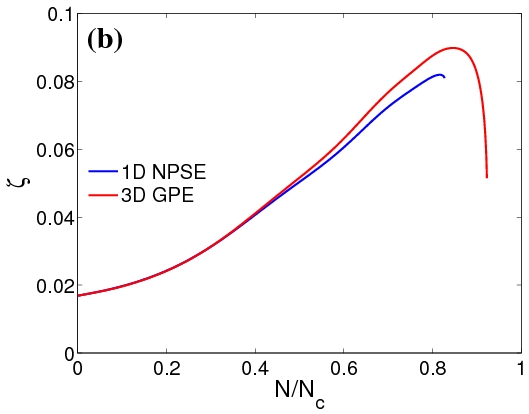} \\
&
\end{tabular}%
\end{center}
\caption{(a) Oscillation asymmetry, $\protect\zeta $ [see the definition of (%
\protect\ref{zeta})], after the first barrier interaction as a function of $%
E $ and $N/N_{c}$, in the 3D setting (as with previous panels, the contour
plot is essentially identical for the quasi-1D case of the NPSE). (b) $%
\protect\zeta $ versus $N/N_{c}$ when $R_{1}=0.5 $. The barrier width is $%
\protect\varepsilon=4.5$ $\mathrm{\protect\mu}$m.}
\label{fig:zeta}
\end{figure}

\begin{figure}[tbp]
\begin{center}
\begin{tabular}{cc}
\includegraphics[bb=0 0 1200 900,width=.51\textwidth]{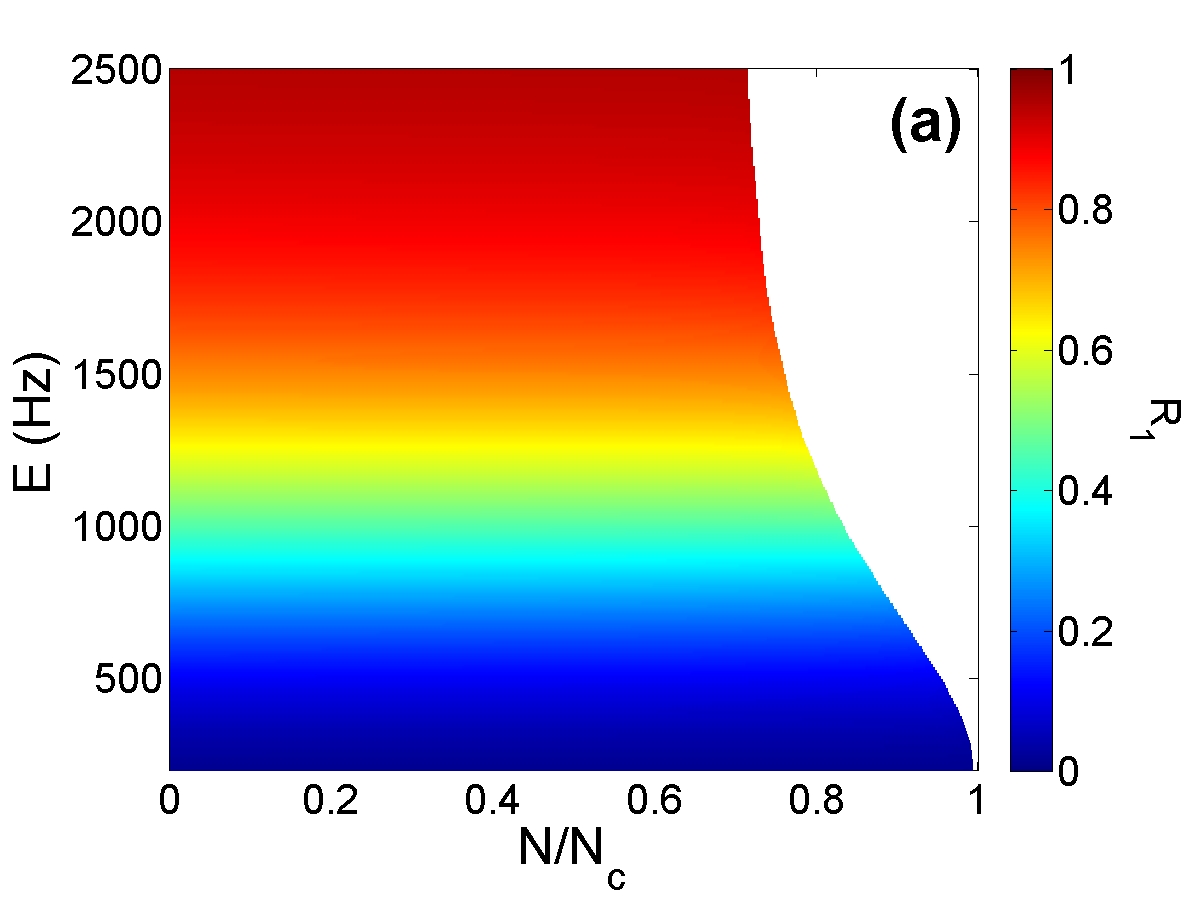} & %
\includegraphics[bb=0 0 1200 900,width=.51\textwidth]{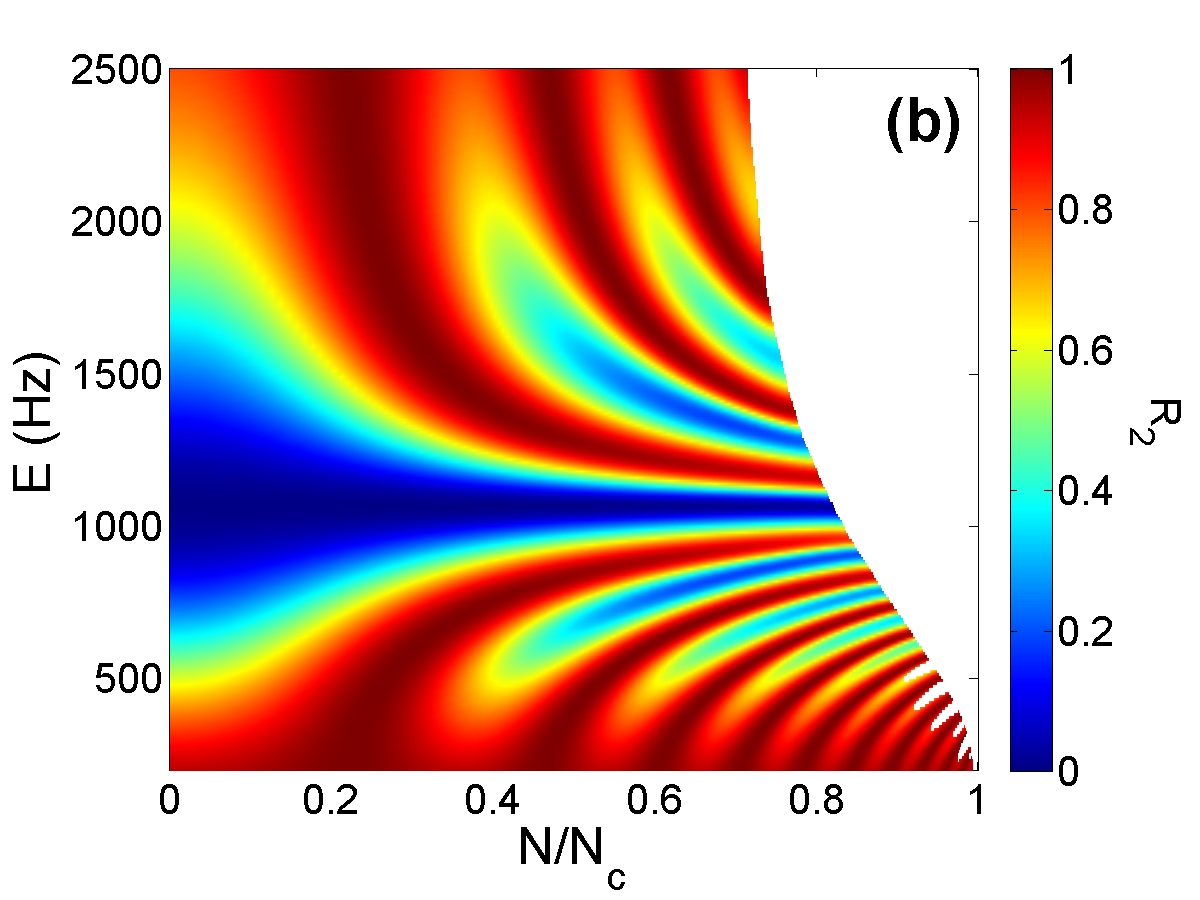} \\
\multicolumn{2}{c}{\includegraphics[bb=0 0 1200 900,width=.51\textwidth]{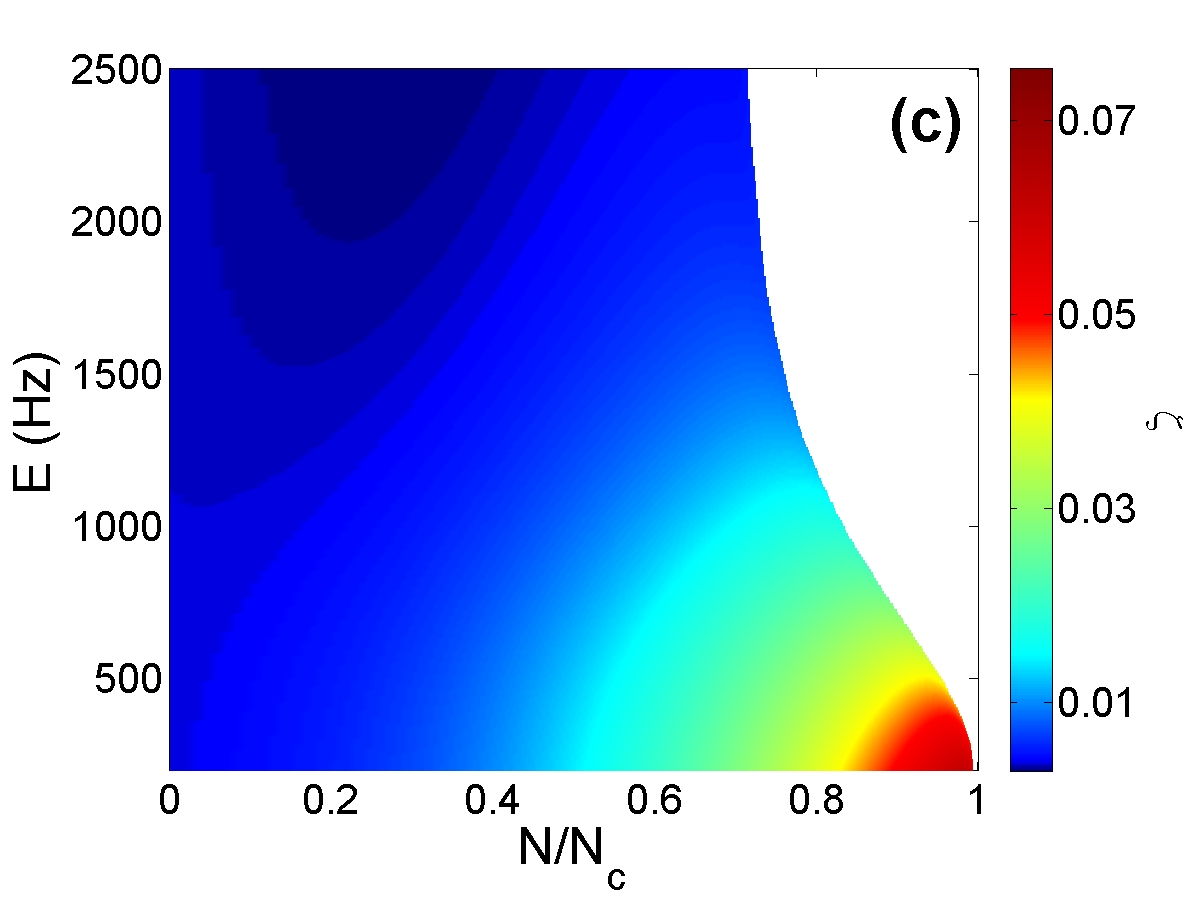}}
\end{tabular}%
\end{center}
\caption{First reflection coefficient, $R_{1}$ (a), second reflection
coefficient $R_{2}$ (b), and the oscillation asymmetry $\protect\zeta$ (c)
vs.~$E$ and $N/N_{c}$, in the quasi-1D setting of the NPSE when $\protect%
\varepsilon=1$ $\mathrm{\protect\mu }$m}
\label{fig:1micron}
\end{figure}

To assess the impact of the width of the barrier on the above results, we
have performed a complementary study for the 1D NPSE, considering a narrower
barrier whose width is $\varepsilon =1$ $\mathrm{\mu }$m, a value much less
than the axial width of the soliton. Figure \ref{fig:1micron} shows the
dependence of $R_{1}$ (a), $R_{2}$ (b) and $\zeta$ (c) with respect to $N$
and $E$ as before. The qualitative nature of the dependence of these
quantities does not seem to change in comparison to the case of $%
\varepsilon=4.5$ $\mu $m. Nevertheless, there is a quantitative shift of the
principal features towards higher values of the barrier height $E$, as
expected.

\begin{figure}[tbp]
\begin{center}
\begin{tabular}{cc}
\includegraphics[bb=0 0 1200 900,width=.51\textwidth]{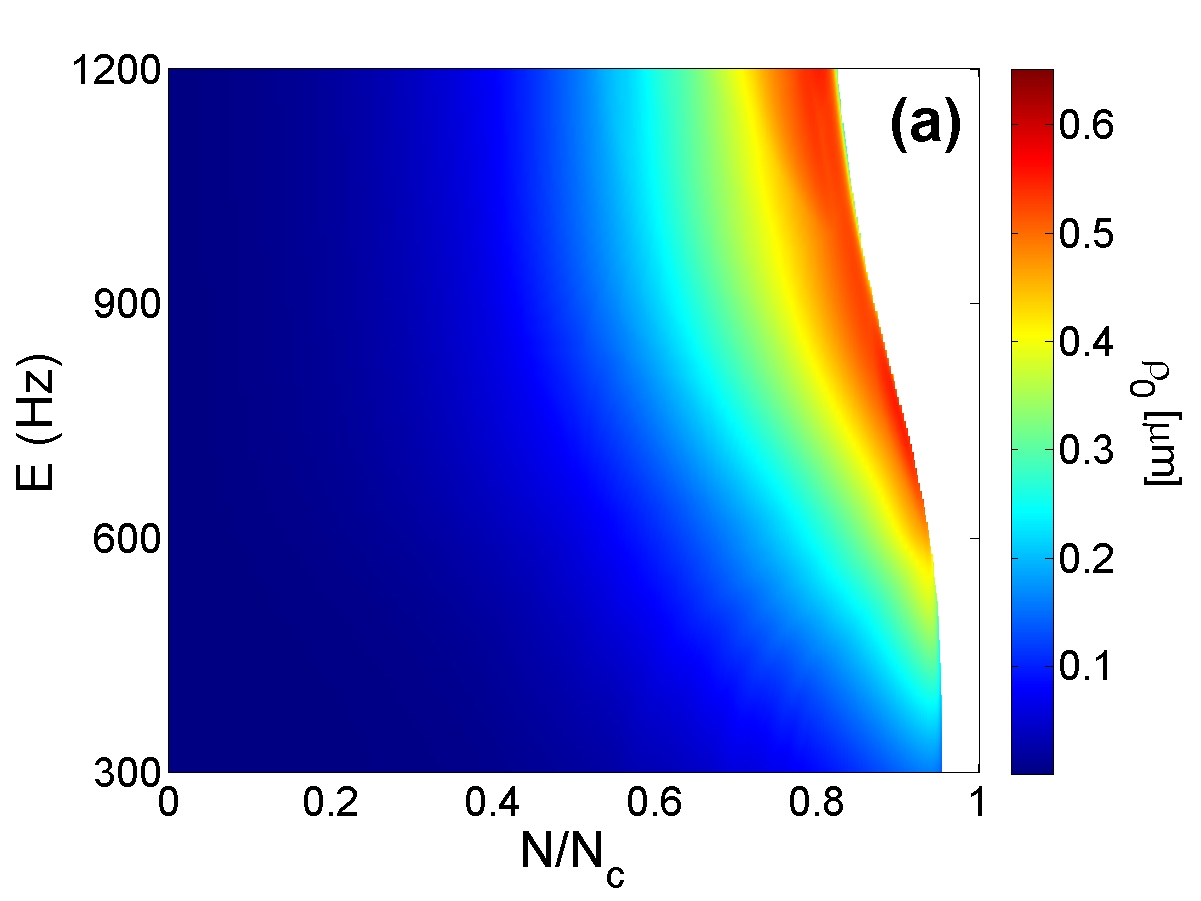} & %
\includegraphics[bb=0 0 541 414,width=.45\textwidth]{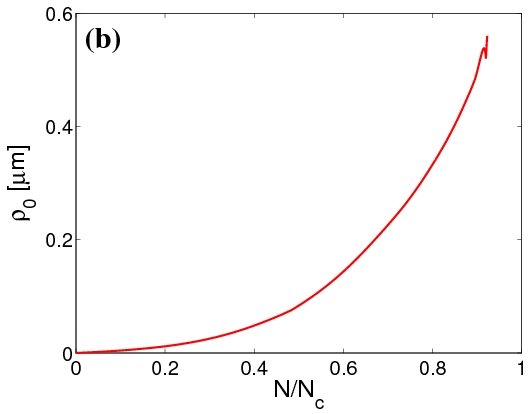}%
\end{tabular}%
\end{center}
\caption{(a) Amplitude $\protect\rho _{0}$ of radial oscillations in the 3D
setting excited by the first interaction with the barrier, as a function of $%
E$ and $N/N_{c}$. In (b), the value of $\protect\rho _{0}$ is plotted versus
$N/N_{c}$, while $R_{1}=0.5$ is fixed. The barrier width is $\protect%
\varepsilon=4.5$ $\mathrm{\protect\mu}$m.}
\label{fig:rho0}
\end{figure}

A basic feature of the 3D dynamics, which may be naturally expected, and is
indeed produced by the 3D GPE, is the excitation of radial oscillations
after splitting. This feature is, by construction, not incorporated in the
1D GPE equation, where the transverse shape is assumed to be
\textquotedblleft frozen". On the other hand, it is incorporated in a simple
(yet reasonably accurate) way in the quasi-1D description of the NPSE,
through the assumption of a space- and time-dependent width of the
transverse direction of the ground state which is directly controlled
(according to the Euler-Lagrange equations) by the longitudinal wave
function. This feature is partially responsible for the good agreement
between the quasi-1D NPSE and the fully 3D GPE, as observed above. We
quantify the radial vibrations in the 3D case by evaluating a measure of
their amplitude

\begin{equation}
\bar{\rho} (t)=\frac{2\pi }{N}\int_{0}^{\infty }\rho d\rho \int_{-\infty
}^{+\infty }dz\,\rho |\psi |^{2},\qquad
\end{equation}%
and defining $\rho_{0}$ as the maximum value of $\bar{\rho}$ (over time for
a given set of parameters). Figure \ref{fig:rho0}(a) shows the dependence of
$\rho _{0}$ on $N/N_{c}$ and $E$, together with its value at the barrier
height $E_{\mathrm{sp}}$ corresponding to $R_{1}=0.5$ (b). It is clearly
observed that with the increase of $N/N_{c}$ and $E$, the excitation of
transverse oscillations becomes very significant, with the amplitude
attaining values $\simeq $ $0.5$ $\mathrm{\mu }$m. The same conclusions are
suggested by the right panel for the case of even splitting. Hence, it can
be inferred that the role of higher-dimensionality (captured qualitatively
within our quasi-1D NPSE approach and properly incorporated in the fully 3D
setting) is of particular relevance for $N$'s close to $N_c$ (a case of
central interest to ongoing experiments).

\begin{figure}[tbp]
\begin{center}
\begin{tabular}{cc}
\multicolumn{2}{c}{\includegraphics[bb=0 0 541 414,width=.45\textwidth]{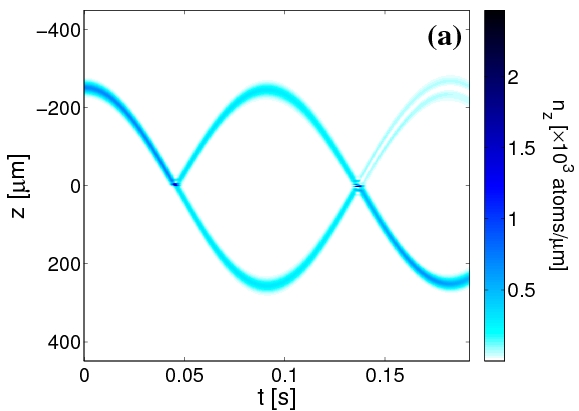}} \\
\includegraphics[bb=0 0 541 414,width=.45\textwidth]{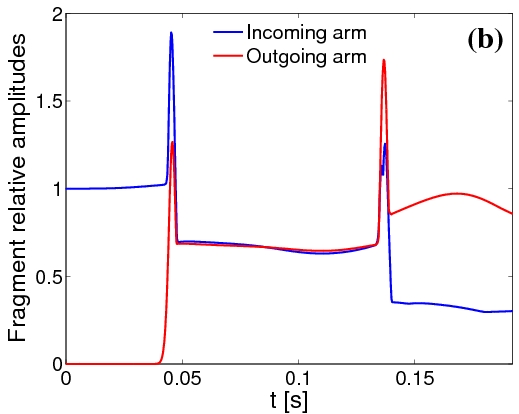} & %
\includegraphics[bb=0 0 541 414,width=.45\textwidth]{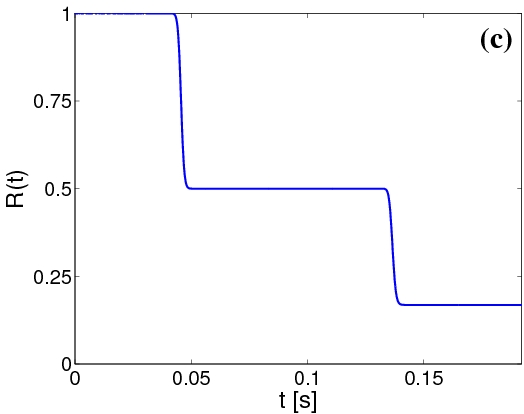} \\
&
\end{tabular}%
\end{center}
\caption{Interaction with a barrier of width $\protect\varepsilon=4.5$ $%
\mathrm{\protect\mu}$m giving $R_{1}=0.5$ and small $R_{2} $ (i.e., even
splitting of the incident soliton followed by the recombination of the
fragments into a nearly single transmitted one) in the 3D GPE (\protect\ref%
{eq:dyn3D}). (a) Integrated density plots. (b) The amplitude of the
fragments divided by the amplitude of the initial soliton (c) The
time-dependent reflection coefficient. The present results were obtained
with $N\approx 0.17N_{c}$ and $E=696.95$ Hz. The smallest value of the
second reflection coefficient is $R_{2}=0.17$. The outcome for the quasi-1D
NPSE is essentially similar.}
\label{fig:R0sim}
\end{figure}

\begin{figure}[tbp]
\begin{center}
\begin{tabular}{cc}
\multicolumn{2}{c}{\includegraphics[bb=0 0 541 414,width=.45\textwidth]{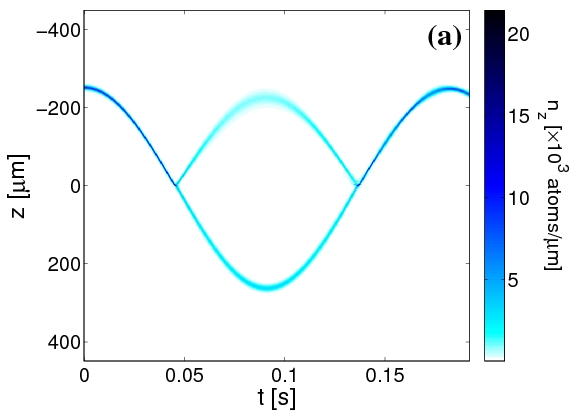}} \\
\includegraphics[bb=0 0 541 414,width=.45\textwidth]{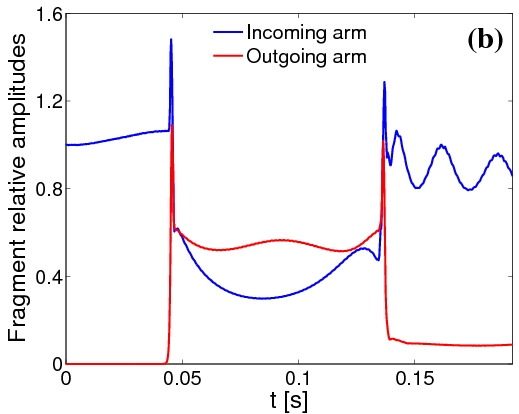} & %
\includegraphics[bb=0 0 541 414,width=.45\textwidth]{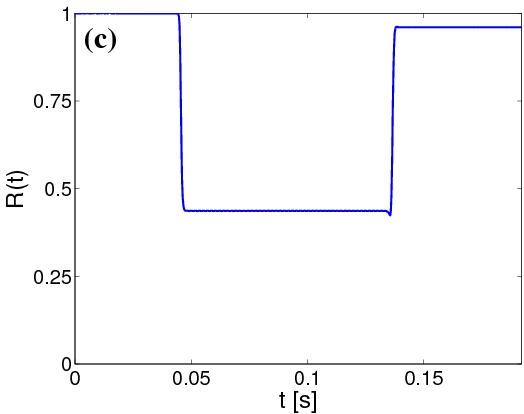} \\
&
\end{tabular}%
\end{center}
\caption{The same as in Figure \protect\ref{fig:R0sim}, but for the case of $%
R_{2}$ close to $1$, i.e., the recombined soliton returning in the incoming
arm. $N\approx 0.88N_{c}$ and $E=686$ Hz, resulting in $R_{1}=0.44$ and $%
R_{2}=0.96$.}
\label{fig:R1sim}
\end{figure}

Figures \ref{fig:R0sim} and \ref{fig:R1sim} show outcomes of simulations
corresponding, respectively, to small and large values of $R_{2}$, in the 3D
setting (results for the quasi-1D NPSE\ are quantitatively similar). The
figures display the longitudinal ($z$-dependent) density, resulting from the
integration of the density in the transverse plane (denoted as $n_z$), along
with amplitudes of the fragments and the time-dependent reflection
coefficients, which exhibit jumps upon the interaction of the one (during
the first event) or of the two (during the second event) solitary wave(s)
with the barrier. As these figures show, the amplitudes of the fragments can
be different even when the numbers of atoms in them are equal. Naturally,
the width is larger for the fragment with the lower peak density.
Furthermore, in addition to the two limiting cases, our results suggest that
one can manipulate the parameters (such as $E$, $N$, etc.) controlling the
interaction of the incident soliton with the barrier to produce any desired
outcome within a wide range in the sequence of two collisions. While there
appear to be two waves in Figure \ref{fig:R0sim} for $z<0$, after the second
soliton-barrier interaction, this feature can be controlled (and avoided)
through the use of a narrower barrier (results not shown here).

\subsection{Effect of changing the initial position of the soliton}

In this subsection the number of atoms is fixed to $N=0.8N_{c}$, which is a
typical, experimentally reachable value where the 3D nature of the
condensate is crucial to the observed dynamics. Hence, we expect the 3D GPE
to be a suitable model, and the quasi-1D NPSE to be a good qualitative (and
reasonable quantitative) approximation. We also expect transverse
excitations to be present in the dynamics.

\begin{figure}[tbp]
\begin{center}
\begin{tabular}{cc}
\includegraphics[bb=0 0 541 414,width=.45\textwidth]{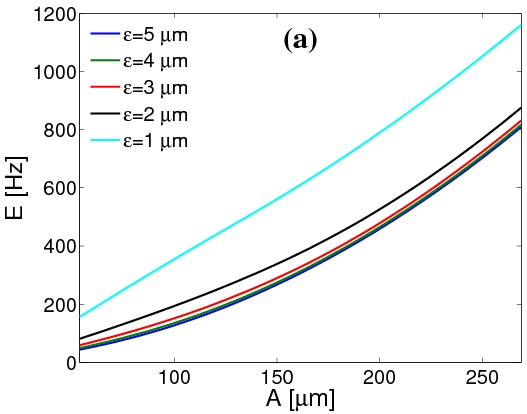} & %
\includegraphics[bb=0 0 541 414,width=.45\textwidth]{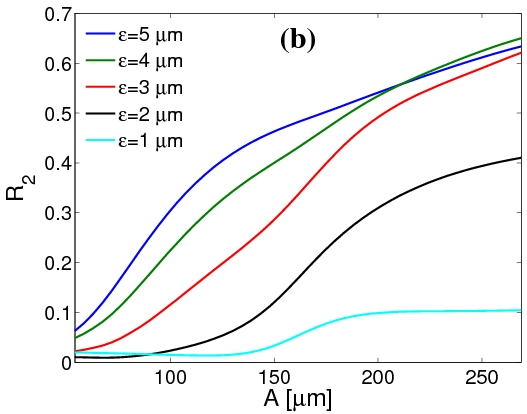} \\
\includegraphics[bb=0 0 541 414,width=.45\textwidth]{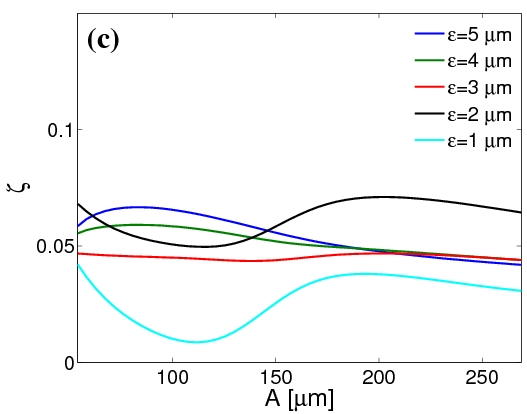} & %
\includegraphics[bb=0 0 541 414,width=.45\textwidth]{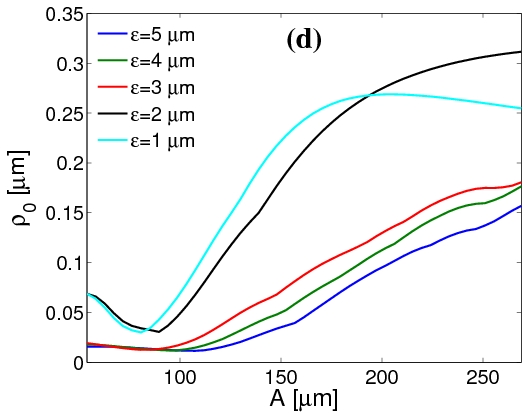} \\
&
\end{tabular}%
\end{center}
\caption{Dependencies of characteristics of the interaction on the initial
displacement of the incident soliton, $A$, for $N=0.8N_{c}$. (a) The barrier
height, $E$, necessary for 50\% splitting after the first collision. For the
following plots, $E = E_{sp}$: (b) The secondary reflection coefficient, $%
R_{2} $; (c) The asymmetry coefficient $\protect\zeta $, defined as per (%
\protect\ref{zeta}); and (d) Amplitude of the radial oscillations, $\protect%
\rho _{0}$. The displayed results are produced by the 3D GPE (\protect\ref%
{eq:dyn3D}).}
\label{fig:Esp}
\end{figure}

We display here the dependence on initial displacement $A$ from which the
soliton is launched with zero velocity (a parameter which effectively
characterizes the
kinetic energy of the incident soliton). For the sake of completeness, we do
this for different barrier widths in (\ref{eq:potential2D}), $\varepsilon
=1,2,3,4$ and $5$ $\mathrm{\mu }$m, in the 3D setting. Figure~\ref{fig:Esp}
shows the dependence on $A$ for the barrier height $E_{\mathrm{sp}}$ giving
even splitting, the second reflection coefficient ($R_{2}$), the asymmetry
parameter (\ref{zeta}), and the radial oscillation amplitude, $\rho _{0}$.
As expected, Figure~\ref{fig:Esp}(a) shows that $E_{\mathrm{sp}}$ increases
with increasing $A$. Under these conditions of strong nonlinearity ($%
N=0.8N_{c}$), however, $E_{\mathrm{sp}}$ depends only weakly on the barrier
width $\varepsilon $, with the exception of the narrowest barrier. Figure~%
\ref{fig:Esp}(b) shows that $R_{2}$ has the least sensitivity to $A$ for the
narrowest barrier with the required height being larger for smaller width,
while $R_{2}$ is larger (as may be expected intuitively) for a larger width.
An increase in $A$ generally leads to only a weak modification of the
asymmetry factor $\zeta $, which is chiefly decreasing with $A$ for wider
barriers, presumably due to the larger speed of the soliton impinging upon
the barrier, and correspondingly smaller interaction times. For narrow
barriers, the dependence on $A$ is more pronounced and non-monotonic. Figure~%
\ref{fig:Esp}(d) shows that transverse excitation is stronger for more
energetic solitons and narrower barriers, where the interaction with the
barrier is more impulsive. We expect transverse excitation to be more
probable when the atomic kinetic energy is comparable to the transverse mode
spacing $\hbar \omega _{\bot }$. This occurs when $A=164\mu $m, which is
reasonably consistent with the sudden increase in $\rho _{0}$ observed for $%
A\simeq 100\mu $m.

We note that the most significant, although small, differences between the
quasi-1D setting of the NPSE and the full 3D setting arise in the case of
narrow barriers. This is intuitively reasonable, as a narrow barrier induces
dynamics on length scales closer to the transverse confinement and hence,
enhances the degree of transverse excitations thus affecting the quality of
the approximation of the 3D behavior by the quasi-1D NPSE. Nevertheless, the
agreement between the two is still in reasonable qualitative agreement.

\subsection{Effects of the phase imprinting}

\begin{figure}[tbp]
\begin{center}
\begin{tabular}{cc}
\includegraphics[bb=0 0 541 414,width=.45\textwidth]{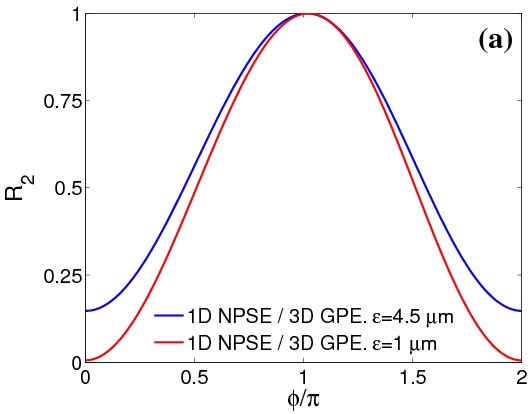} & %
\includegraphics[bb=0 0 541 414,width=.45\textwidth]{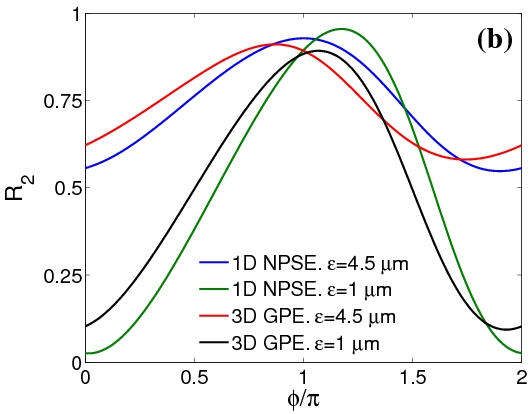}%
\end{tabular}%
\end{center}
\par
. .
\caption{Second reflection coefficient, $R_{2}$, in the case of the initial
50\% splitting, as a function of the phase difference introduced between the
fragments at the time of their largest separation. The panels correspond to
(a) $N\approx 0.06N_{c}$ (near the linear limit, shown for comparison) and
(b) $N\approx 0.8N_{c}$. Notice that in the former case, the results for 1D
and 3D settings coincide for $\protect\varepsilon=4.5$ $\mathrm{\protect\mu}$%
m. In all cases, the initial displacement of the soliton was $A=251$ $%
\mathrm{\protect\mu }$m}
\label{fig:phase1}
\end{figure}

Finally, we briefly analyze the effect of imprinting a phase difference at a
certain time, $t=\pi /\omega _{z} $ onto the initially transmitted fragment,
thereby emulating the operation of an interferometer. The phase difference
is introduced when the fragments are located at the largest distance from
the barrier. Figure \ref{fig:phase1} shows the second reflection
coefficient, $R_{2}$, in the cases of small (for comparison) and large atom
numbers, as obtained from the simulations of the quasi-1D NPSE (\ref%
{eq:dyn1D}) and the 3D GPE (\ref{eq:dyn3D}). It is seen that, in the regimes
of small $N$, the absence of a phase shift provides for a nearly complete
transmission, while a relative phase shift of $\pi $ leads to almost perfect
reflection. The contrast is considerably reduced for strong nonlinearity and
sufficiently wide barriers, as shown in the right panel of Figure \ref%
{fig:phase1}, although a wider range of transmittivities/reflectivities is
accessible for narrower barriers. In that light, although highly nonlinear
waves (such as solitons) may be deemed less useful for interferometric
purposes, their relevance may be (at least partially) restored in the
context of suitably narrow barriers.

\section{Conclusion}

We have explored the collision of matter-wave solitons with a central
barrier of variable width and height inserted into a shallow
harmonic-oscillator axial trap, and the subsequent collision of the split
fragments upon their return to the defect. This configuration forms the
basis of a soliton interferometer. The non-monotonic variation of $R_2$ with
$N$ and $E$, evident for strong nonlinearity in Figures \ref{fig:R2} and \ref%
{fig:1micron}, is detrimental to the sensitivity of an atom interferometer.
Figure \ref{fig:phase1}, however, indicates that the sensitivity to phase
variation can be regained by smaller nonlinearity and that the adverse
effect of strong nonlinearity can be partially mitigated by a narrow
barrier. Narrow in this context, evidently means in comparison to the axial
size of the soliton.

Our analysis accounted for relatively large atom numbers, which are not much
smaller than the collapse threshold. This strong nonlinearity necessitates a
full 3D solution, in contrast to the 1D GPE setting studied in previous
works. The analysis was carried out, in parallel, in the framework of the
quasi-1D NPSE and the full 3D GPE, indicating a good qualitative and even
reasonable quantitative (at least not too close to $N=N_c$) agreement
between the two. A detailed computational map of the ensuing phenomenology
has been generated as a function of number of atoms in the soliton, its
initial distance from the barrier (which determines the collision velocity),
and the height and width of the splitting barrier. Additionally, the effect
of a phase shift imposed on the fragments at the moment of the largest
separation was also examined.

A number of general conclusions, obtained in the framework of the quasi-1D
and full 3D settings, and their similarities and differences have been
reported. While the results are similar between these two cases, they are
essentially different from those generated by the 1D cubic GPE, which was
used previously. The quasi-1D NPSE and 3D GPE produce similar values for the
first reflectivities, systematically increasing with increasing barrier
height $E$ and weakly decreasing with increasing number of atoms $N$. The
second reflectivity (corresponding to the collision of the fragments after
the first interaction with the barrier) oscillates strongly from complete
reflection to high transmission as function of barrier height (and atom
number). These oscillations are a fundamentally nonlinear effect, most
pronounced for large $N$. The excitation of the transverse breathing mode
was also characterized in the framework of the 3D GPE. This effect was found
to become progressively more significant as the critical number of atoms,
corresponding to the onset of the collapse, was approached. Phase imprinting
was found to play a critical role in the outcome of the second collision,
especially for small nonlinearity, modifying it between nearly complete
transmission and full reflection. The highly nonlinear realm seems less
sensitive to such variations, but the sensitivity may be restored for
narrower barriers.

It is important to corroborate these findings experimentally. Regarding
further theoretical analysis, a challenging problem is to study deviations
of the results from the mean-field approximation in the 3D geometry as a
function of the atom number $N$, in analogy to the recently emergent studies
in the 1D setting \cite{martin,Lev,France}. An additional issue of interest
is the use of potential wells, rather than barriers, for which more complex
phenomenology may be expected in the 3D setting \cite{pantofl1,pantofl2}.
Lastly, extending such considerations to dark solitons in condensates with
the self-repulsive nonlinearity \cite{fotis}, and to multi-component~systems
\cite{Denmark,usnow} are also compelling topics for further investigation.

\ack

This work was supported, in a part, by grant No. 2010239 from the Binational
(US-Israel) Science Foundation. J.C. acknowledges financial support from the
MICINN project FIS2008-04848. The work at Rice was supported by the NSF
(PHY-1102515), ONR, the Norman Hackerman Advanced Research Program of Texas,
and the Welch Foundation (C-1133). PGK acknoweldges support from the US-NSF
through grant DMS-0806762 and from the Alexander von Humboldt Foundation. We
are indebted to Faustino Palmero for setting up the HPC cluster where the
simulations were performed.

\end{document}